\documentclass[adp,fleqn]{w-art}
\usepackage{times}
\usepackage{w-thm}
\usepackage[]{graphicx}

\usepackage{hyperref}
\usepackage{amsfonts}

\providecommand{\eprint}[1]{\href{http://arxiv.org/abs/#1}{{\tt [arXiv:#1]}}}

\providecommand\apj{ApJ}                 
\providecommand\apjs{ApJSupp}                 
\providecommand\aap{A\&A}            
\providecommand\mnras{MNRAS}
\providecommand\nat{Nature}
\providecommand\cqg{CQG}

\providecommand\prd{Phys.~Rev.~D}
\providecommand\physrep{Phys. Rep.}

\providecommand\hGpc{\mbox{$h^{-1}$ Gpc}}

\newcommand\rC{R_C}

\begin{document}
\DOIsuffix{theDOIsuffix}
\Volume{16}
\Issue{1}
\Copyrightissue{01}
\Month{01}
\Year{2007}
\pagespan{1}{} 
\Receiveddate{}
\Reviseddate{}
\Accepteddate{}
\Dateposted{}



\title{Some spaces are more equal than others}


\author[B.F. Roukema]{Boudewijn F. Roukema\inst{1,}%
}
\address[\inst{1}]{Toru\'n Centre for Astronomy, Nicolaus Copernicus University,
ul. Gagarina 11, 87-100 Toru\'n, Poland}
\begin{abstract}
It has generally been thought that in perturbed
Friedmann-Lema\^{\i}tre-Robertson-Walker models of the Universe, global
topology should not have any feedback effects on dynamics.  However, a
weak-field limit heuristical argument, assuming a finite particle
horizon for the transmission of gravitational signals, shows that a
residual acceleration effect can occur.  The nature of this effect
differs algebraically between different constant curvature
3-manifolds.  This potentially provides a selection mechanism
for the 3-manifold of comoving space.
\end{abstract}
\maketitle                   






\section{Cosmic topology and dynamics}

The geometrical and global topological freedom allowed by constant
curvature Riemannian 3-manifolds as a model of the comoving spatial
section of a relativistic universe
\cite{deSitt17,Fried23,Fried24,Lemaitre31ell,Rob35} imply that a
Friedmann-Lema\^{\i}tre-Robertson-Walker (FLRW) model may be finite in
total comoving spatial volume for any of the three curvatures:
negative, zero or positive. This resolves the historical dilemma in
cosmology according to which two physically desirable qualities of
space seemed to be contradictory: a finite universe without boundaries
seemed to be a self-contradiction. Riemannian geometry gave a solid
mathematical foundation to the resolution of this apparent conflict.
Even before the relavistic era, Riemannian 3-manifolds of curvature
and/or topology different to that of simply-connected Euclidean space
were proposed as models of space by Karl Schwarzschild
\cite{Schw00,Schw98}.  Skipping forward, the start of the ``modern''
epoch of active cosmic topology research can probably be dated back to
the two 1993 papers by Starobinsky \cite{Star93} and Stevens et
al. \cite{Stevens93}.  It is a great pleasure to have
Alexei Starobinsky here with us in this meeting.  For reviews of the
historical and modern theoretical and observational aspects of cosmic
topology, see refs \cite{LaLu95,Lum98,Stark98,LR99,BR99,RG04}.

Although, in general, these 3-manifolds are not vector spaces,
Grassmann's pioneering work in linear algebra
\cite{Grass44A1,Grass62A2} nevertheless provides useful tools (as it
does in nearly all of modern science), since working in the covering
spaces $ \mathbb{H}^3 $, $\mathbb{R}^3$, and $ S^3 $, for negative,
zero, and positive curvature, respectively, is useful for many
observational and theoretical purposes.  The covering space
$\mathbb{R}^3$ is a vector space, and calculations made when the
covering space is the hypersphere $S^3$ are very straightforward to
program when $S^3$ is embedded in $\mathbb{R}^4$. Both the analytical
and numerical calculations referred to below for the spherical
well-proportioned spaces specifically use this latter
technique---thanks to Hermann Grassmann.

What was long taken as self-evident in cosmic topology research was
the inference that since the Einstein field equations are local, there
is no way that the global topology of the spatial section of an FLRW
universe could have an effect on that universe's dynamics. The only
known link between topology and dynamics was through curvature,
since the three different curvatures allow three different families of 3-manifolds.

\section{Residual gravity}
However, it was shown heuristically in ref~\cite{RBBSJ06} that in a
multiply-connected universe containing a density perturbation, the
effects of the distant copies of the perturbation in the covering
space are not exactly symmetrical on a massless test particle, leading
to a non-zero residual acceleration effect.  Consider Fig.~3 in
ref~\cite{RBBSJ06}. In a flat, multiply-connected model of comoving
in-diameter $L$, the perturbation is modelled as a point-sized massive
object, and the gravitational pull on a massless test particle at a
small distance $x$ is estimated in the weak-field limit,
i.e. Newtonian gravity within a finite particle horizon is used. In
the covering space $\mathbb{R}^3$ along a given holonomy
transformation axis, one of the two adjacent copies (at $L-x$ and
$L+x$) of the massive object is slightly closer to the test particle
than the other. Hence, at the position of the massless test particle,
after removing the gravitational pull towards the ``original'' copy of
the massive object at distance $x$, there is a residual gravitational
force that pulls the massless test particle towards the slightly
closer of the two topological images of the ``original'' massive 
object.\footnote{Physically, there is no difference between an ``original'' 
and a ``copy'' of the object---these are two images in the covering space.
See the review papers cited above for a fuller introduction.}
A Taylor expansion in
$x/L$ shows that the residual acceleration is dominated by the first-order term
\begin{equation}
\ddot{x} = \frac{4Gm}{L^3} x
\label{e-T1}
\end{equation}
where $G$ is the gravitational constant and $m$ is the mass of the massive object.

This example does not constitute a full relativistic calculation, but it is
difficult to see how the effect could be avoided in a perturbed FLRW model,
or in an hypothetical exact solution for an almost FLRW model. A universe
which is small and homogeneous except for one positive density fluctuation 
is seen by an observer to be slightly anisotropic unless the
observer is located at the centre of the fluctuation, which can be assumed
to be spherically symmetric. This anisotropy concerns the gravitational 
potential seen from different directions, as it is transmitted to the observer
by gravitational waves. It is difficult to avoid the conclusion that 
the dynamics of a perturbed FLRW model can be affected by global topology,
even if the effect at the present epoch is likely to be small.

\subsection{Stabilisation towards equal fundamental lengths}
In a right-angled 3-torus model, i.e. $T^3 \equiv
\mathbb{R}^3/\mathbb{Z}^3$ where the fundamental domain is a
rectangular prism, if the three side-lengths of the prism $L_i$ are
very unequal to one another, then the $L_i^{-3}$ factor will cause the
shortest length to induce the strongest residual acceleration. As
suggested in Sect.~3.2.3 of \cite{RBBSJ06}, this may cause the shorter
length of the fundamental domain to expand faster than the other two
directions, tending towards equality of the three side-lengths.  The
effects would equalise when the side-lengths equalise. If the model
initially has an isotropic scale factor, then it will become
anisotropic in the sense that the scale factors in different
directions will become slightly unequal, and will remain so until
equal side-lengths are obtained.

\subsection{$T^3$ is both well-proportioned and well-balanced}
However, at the equilibrium state of equal side-lengths of a $T^3$ model,
there residual acceleration is different to that given in Eq.~(\ref{e-T1}).
As shown algebraically and numerically in Sect.~3.1 of ref~\cite{RR09}, 
the residual accelerations 
induced by a massive object 
in an exactly regular $T^3$ model 
cancel down to the third order in $\epsilon_x,
\epsilon_y, \epsilon_z$, the distances to the test particle as fractions
of the respective fundamental lengths. Hence, not only is a regular 
$T^3$ model an equilibrium state in the sense that the residual accelerations
in the three directions will tend to equalise, but the residual acceleration
as a vector ($T^3$ is a flat space) will also drop sharply in amplitude
as this equalisation is approached.

Spaces that have approximately equal fundamental lengths have 
been termed ``well-proportioned'' \cite{WeeksWellProp04}. It is now clear
that (regular) $T^3$ is not only well-proportioned:
it is also well-balanced.

\section{Well-proportioned spherical FLRW models with a perturbation}

Well-proportioned spherical FLRW models also exist. While well-proportionality
 might seem to be a subjective, aesthetic criterion for
a preferred model of the Universe, the lack of structure on scales
above $\sim 10{\hGpc}$ in the Wilkinson Microwave Anisotropy Probe
(WMAP) sky maps, predicted as a sign of cosmic topology based on 
the low-resolution COBE maps \cite{Star93,Stevens93},
 is best explained by a well-proportioned model 
\cite{WeeksWellProp04}. Among
these, the Poincar\'e dodecahedral space model, $S^3/I^*$, has become a
particularly good (though disputed)
candidate \cite{WMAPSpergel,LumNat03,RLCMB04,Aurich2005a,Aurich2005b,Gundermann2005,KeyCSS06,NJ07,Caillerie07,LR08,RBSG08,RBG08}.

Are the well-proportioned spherical spaces, $S^3/T^*$, $S^3/O^*$, and
$S^3/I^*$, also well-balanced in the sense of the residual gravity
effect? As shown by Grassmann \cite{Grass44A1} over a century and a
half ago, $\mathbb{R}^4$ exists as a self-consistent mathematical
object---a four-dimensional vector space---and on a modern computer,
calculations in $\mathbb{R}^4$ require only a minor change in computer
code compared to those in $\mathbb{R}^3$. By embedding $S^3$ in $\mathbb{R}^4$,
both algebraic and numerical calculations of the dominant terms of the
residual gravity effect are rendered tractable for these spaces, as
presented in ref~\cite{RR09}.

The result is that all three of these spaces are indeed well-balanced.
The linear term in the Taylor expansion of the residual gravity effect
cancels in all three cases.  However, the Poincar\'e space $S^3/I^*$,
the space that has been selected by empirical arguments, is even
better balanced than the other spaces. Not only does the linear term
cancel, but the third-order term also cancels, leaving an expression
dominated by the fifth order.  This fifth-order term can be written as
a vector in $\mathbb{R}^4$
\begin{eqnarray}
 \ddot{\mathbf{r}} &=&
\frac{
12 \sqrt{2} \left(297 \sqrt{5} + 655\right)
}{125 \sqrt{5-\sqrt{5}}}
\left(\frac{r}{\rC}\right)^5
 \nonumber \\  && 
\Big\{
\big[70\,y^4
+(42\,\sqrt{5}+70)\,x^2\,y^2
-(14\,\sqrt{5}+70)\,y^2    
+(21\,\sqrt{5}-7)\,x^4
-28\,\sqrt{5}\,x^2          \nonumber \\  && 
+7\,\sqrt{5}+5\big] \;x,    \nonumber \\  && 
\big[70\, z^4
+(42\,\sqrt{5}+70)\,y^2\,z^2
-(14\,\sqrt{5}+70)\,z^2
+(21 \,\sqrt{5}-7)\,y^4
-28\,\sqrt{5}\,y^2       \nonumber \\  && 
+7\,\sqrt{5}+5\big] \;y,     \nonumber \\  && 
\big[
70\,x^4
+(42\,\sqrt{5}+70) \,x^2\,z^2
-(14\,\sqrt{5}+70)\,x^2
+(21\,\sqrt{5}-7)\,z^4
-28\,\sqrt{5}\,z^2         \nonumber \\  && 
+7\,\sqrt{5}+5\big] \;z ,     \nonumber \\  && 
0
\Big\} , 
\label{e-residgrav-Poincare-exact}
\end{eqnarray}
where the massive particle is at $(0,0,0,1)$, the
curvature radius is $\rC$, the 
nearby test particle is at
$\mathbf{p} := [\sin(r/\rC) x, \sin(r/\rC) y , \sin(r/\rC) z, \cos(r/\rC)]$, 
$x^2+y^2+z^2=1$, and the time component is not represented geometrically
(cf. Eq.~(21), \cite{RR09}). 
The spatial direction of the residual acceleration in
Eq.~(\ref{e-residgrav-Poincare-exact}) appears to be tangent to the
observer at $(0,0,0,1)$, but this is only because terms higher than fifth
order have been dropped. The dominant term in the $w$ direction is a 
sixth-order term. Including this term shows that the residual acceleration 
is indeed in the tangent 3-plane to the 3-sphere at $\mathbf{p}$, and not
in the tangent 3-plane to the observer.

Thus, regular $T^3$, octahedral space $S^3/T^*$, and truncated cube
space $S^3/O^*$ all have residual gravity effects sufficiently
balanced that they are dominated by the third-order term, while the
Poincar\'e space is even better balanced, with the fifth-order term
dominating.  Hence, not only does the residual gravity effect show
that global topology can effect the dynamics of a perturbed FLRW
universe, but the strength of the effect differs between different
spatial sections, and one space has an effect that cancels even more
perfectly than in the other spaces: {\em some spaces are more equal
  than others} \cite{Orwell45}.

\section{Conjecture: did residual gravity select the Poincar\'e space?}

The residual gravity effect provides a simple mechanism by which
the global topology of the comoving spatial section of the Universe
can have a feedback effect on the dynamics of how the Universe itself
expands. While only very elementary calculations have been performed so
far, the initial results are tantalising. Not only does the effect 
seem to be a stabilising effect towards equal side-lengths in a $T^3$
model, but it also seems that the effect selects
out the space that is preferred {\em empirically} by several groups based on the 
WMAP data---the Poincar\'e dodecahedral space, $S^3/I^*$---as being
better balanced than other spaces. Could this effect 
have provided a selection criterion during the quantum epoch of the
Universe?

\begin{acknowledgement}
  The author thanks the organisers for a meeting that was very enjoyable
and scientifically productive.
\end{acknowledgement}



\begin{thebibliography}{10}

\bibitem{Aurich2005a} {Aurich} R.,  {Lustig} S.,    {Steiner} F.,
2005a, \cqg, 22, 3443, \eprint{astro-ph/0504656}.

\bibitem{Aurich2005b} {Aurich} R.,  {Lustig} S.,    {Steiner} F.,
2005b, \cqg, 22, 2061, \eprint{astro-ph/0412569}.

\bibitem{BR99} {Blanl{\oe}il} V.,  {Roukema} B.~F.,  eds, 2000,
``Cosmological Topology in Paris 1998'' Paris: Blanl{\oe}il \& Roukema,
\eprint{astro-ph/0010170}.

\bibitem{Caillerie07} {Caillerie} S.,  {Lachi{\`e}ze-Rey} M.,
{Luminet} J.~.,  {Lehoucq} R., {Riazuelo} A.,    {Weeks} J.,  2007,
\aap, 476, 691, \eprint{0705.0217v2}.

\bibitem{deSitt17} {de Sitter} W.,  1917, \mnras, 78, 3.

\bibitem{Fried23} {Friedmann} A.,  1923, {{\sl Mir kak prostranstvo
i vremya} (The Universe as Space and Time)}.  Leningrad: Academia.

\bibitem{Fried24} {Friedmann} A.,  1924, Zeitschr. f\"ur Phys.,
21, 326.

\bibitem{Grass44A1} {Grassmann} H.~G.,  1844, {{\sl Die lineare
Ausdehnungslehre}}.  Leipzig: Wiegand.

\bibitem{Grass62A2} {Grassmann} H.~G.,  1862, {{\sl Die
Ausdehnungslehre, vollst\"andig und in strenger Form bearbeitet}}.
Berlin: Enslin.

\bibitem{Gundermann2005} {Gundermann} J.,  2005, ArXiv e-prints,
\eprint{astro-ph/0503014}.

\bibitem{KeyCSS06} {Key} J.~S.,  {Cornish} N.~J.,  {Spergel} D.~N.,
{Starkman} G.~D.,  2007, \prd, 75, 084034, \eprint{astro-ph/0604616}.

\bibitem{LaLu95} {Lachi\`eze-Rey} M.,  {Luminet} J.,  1995, \physrep,
254, 135, \eprint{gr-qc/9605010}.

\bibitem{Lemaitre31ell} {Lema{\^i}tre} G.,  1931, \mnras, 91, 490.

\bibitem{LR08} {Lew} B.,  {Roukema} B.~F.,  2008, \aap, 482, 747,
\eprint{0801.1358}.

\bibitem{LR99} {Luminet} J.,  {Roukema} B.~F.,  1999, in NATO
ASIC Proc. 541: Theoretical and Observational Cosmology. Publisher:
Dordrecht: Kluwer, {Topology of the Universe: Theory and Observation}.
p.~117, \eprint{astro-ph/9901364}.

\bibitem{LumNat03} {Luminet} J.,  {Weeks} J.~R.,  {Riazuelo}
A.,  {Lehoucq} R.,    {Uzan} J., 2003, \nat, 425, 593,
\eprint{astro-ph/0310253}.

\bibitem{Lum98} {Luminet} J.-P.,  1998, Acta Cosmologica, XXIV-1,
105, \eprint{gr-qc/9804006}.

\bibitem{NJ07} {Niarchou} A.,  {Jaffe} A.,  2007, Physical Review
Letters, 99, 081302, \eprint{astro-ph/0702436}.

\bibitem{Orwell45} {Orwell} G.,  1945, {Animal Farm: A Fairy Story}.
London: Secker and Warburg.

\bibitem{RG04} {Rebou\c{c}as} M.~J.,  {Gomero} G.~I.,  2004,
Braz. J. Phys., 34, 1358, \eprint{astro-ph/0402324}.

\bibitem{Rob35} {Robertson} H.~P.,  1935, \apj, 82, 284.

\bibitem{RBBSJ06} {Roukema} B.~F.,  {Bajtlik} S.,  {Biesiada}
M.,  {Szaniewska} A., {Jurkiewicz} H.,  2007, \aap, 463, 861,
\eprint{astro-ph/0602159}.

\bibitem{RBG08} {Roukema} B.~F.,  {Buli\'nski} Z.,    {Gaudin} N.~E.,
2008, \aap, 492, 673, \eprint{0807.4260}.

\bibitem{RBSG08} {Roukema} B.~F.,  {Buli\'nski} Z.,  {Szaniewska}
A.,    {Gaudin} N.~E.,  2008, \aap, 486, 55, \eprint{0801.0006}.

\bibitem{RLCMB04} {Roukema} B.~F.,  {Lew} B.,  {Cechowska}
M.,  {Marecki} A.,    {Bajtlik} S., 2004, \aap, 423, 821,
\eprint{astro-ph/0402608}.

\bibitem{RR09} {Roukema} B.~F.,  {R\'o\.za\'nski} P.~T.,  2009, \aap,
502, 27, \eprint{0902.3402}.

\bibitem{Schw00} {Schwarzschild} K.,  1900, Vier.d.Astr.Gess, 35, 337.

\bibitem{WMAPSpergel} {Spergel} D.~N.,  {Verde} L.,  {Peiris}
H.~V.,  {Komatsu} E.,  {Nolta} M.~R., {Bennett} C.~L.,  {Halpern}
M.,  {Hinshaw} G.,  {Jarosik} N.,  {Kogut} A., {Limon} M.,  {Meyer}
S.~S.,  {Page} L.,  {Tucker} G.~S.,  {Weiland} J.~L., {Wollack}.  E.,
{Wright} E.~L.,  2003, \apjs, 148, 175, \eprint{astro-ph/0302209}

\bibitem{Stark98} {Starkman} G.~D.,  1998, \cqg, 15, 2529.

\bibitem{Star93} {Starobinsky} A.~A.,  1993, Journal of Experimental
and Theoretical Physics Letters, 57, 622.

\bibitem{Stevens93} {Stevens} D.,  {Scott} D.,    {Silk} J.,  1993,
Physical Review Letters, 71, 20.

\bibitem{Schw98} {Stewart} J.~M.,  {Stewart} M.~E.,    {Schwarzschild}
K.,  1998, \cqg, 15, 2539.

\bibitem{WeeksWellProp04} {Weeks} J.,  {Luminet} J.-P.,  {Riazuelo} A.,
{Lehoucq} R.,  2004, \mnras, 352, 258, \eprint{astro-ph/0312312}.

\end{thebibliography}
\end{document}